\documentclass[preprint,3p,12pt]{elsarticle}

\usepackage{lineno,hyperref}
\modulolinenumbers[5]

\usepackage{epstopdf}

\usepackage{mathrsfs}
\usepackage{amssymb}
\usepackage{amsfonts}
\usepackage{latexsym}
\usepackage{amsmath}
\usepackage{xcolor}
\usepackage{wrapfig}
\usepackage{floatflt}

\usepackage{graphicx}

\def\lb{\label}

\newcommand{\er}[1]{\textrm{(\ref{#1})}}

\newtheorem{theorem}{\bf Theorem}[section]

\def\a{\alpha}         
\def\b{\beta}

\def\d{\delta}         
\def\D{\Delta}  \def\cF{{\mathcal F}}       \def\mF{{\mathscr F}}
    \def\cG{{\mathcal G}}

 \def\cN{{\mathcal N}}

  \def\cR{{\mathcal R}}       
  \def\cS{{\mathcal S}}

\def\o{\omega}

\def\ve{\varepsilon}           

\def\Z{{\mathbb Z}}       \def\C{{\mathbb C}}

\def\lt{\biggl}                  \def\rt{\biggr}

                 \let\le\leqslant

\def\ss{\subset}                 
\def\pa{\partial}

\def\el2{\ell^{\,2}}             \def\1{1\!\!1}

\def\det{\mathop{\mathrm{det}}\nolimits}
\def\diag{\mathop{\mathrm{diag}}\nolimits}

\def\rank{\mathop{\mathrm{rank}}\limits}

\let\le\leqslant

\newcommand{\ca}{\begin{cases}}
\newcommand{\ac}{\end{cases}}
\newcommand{\ma}{\begin{pmatrix}}
\newcommand{\am}{\end{pmatrix}}
\def\eq{\begin{equation}}
\def\qe{\end{equation}}
\def\[{\begin{equation}}
\def\]{\end{equation}}

\begin{document}

\begin{frontmatter}

\title{Recovery of defects from the information at detectors}
\date{\today}


\author
{Anton A. Kutsenko}

\address{Department of Mathematics, Aarhus University, DK-8000,
Denmark; email: akucenko@gmail.com}

\begin{abstract}
The discrete wave equation in a multidimensional uniform space with
local defects and sources is considered. The characterization of all
possible defect configurations corresponding to given amplitudes of
waves at the receivers (detectors) is provided.
\end{abstract}


\end{frontmatter}


\section{Introduction}

The recovery of defects from the available information about
amplitudes of waves at the detectors is very important in
non-destructive testing in general and appear in areas such as
structural geology inversion, medical imaging, and modeling of
cloaking devices, see discussion in \cite{K}. The popular methods of
solving this problem is based on stochastic inversion. In the
current paper we describe analytically the set of all possible
defects corresponding to given amplitudes of waves at the detectors.

\begin{figure}[h]
{\begin{minipage}[h]{0.49\linewidth}
\center{\includegraphics[width=1\linewidth]{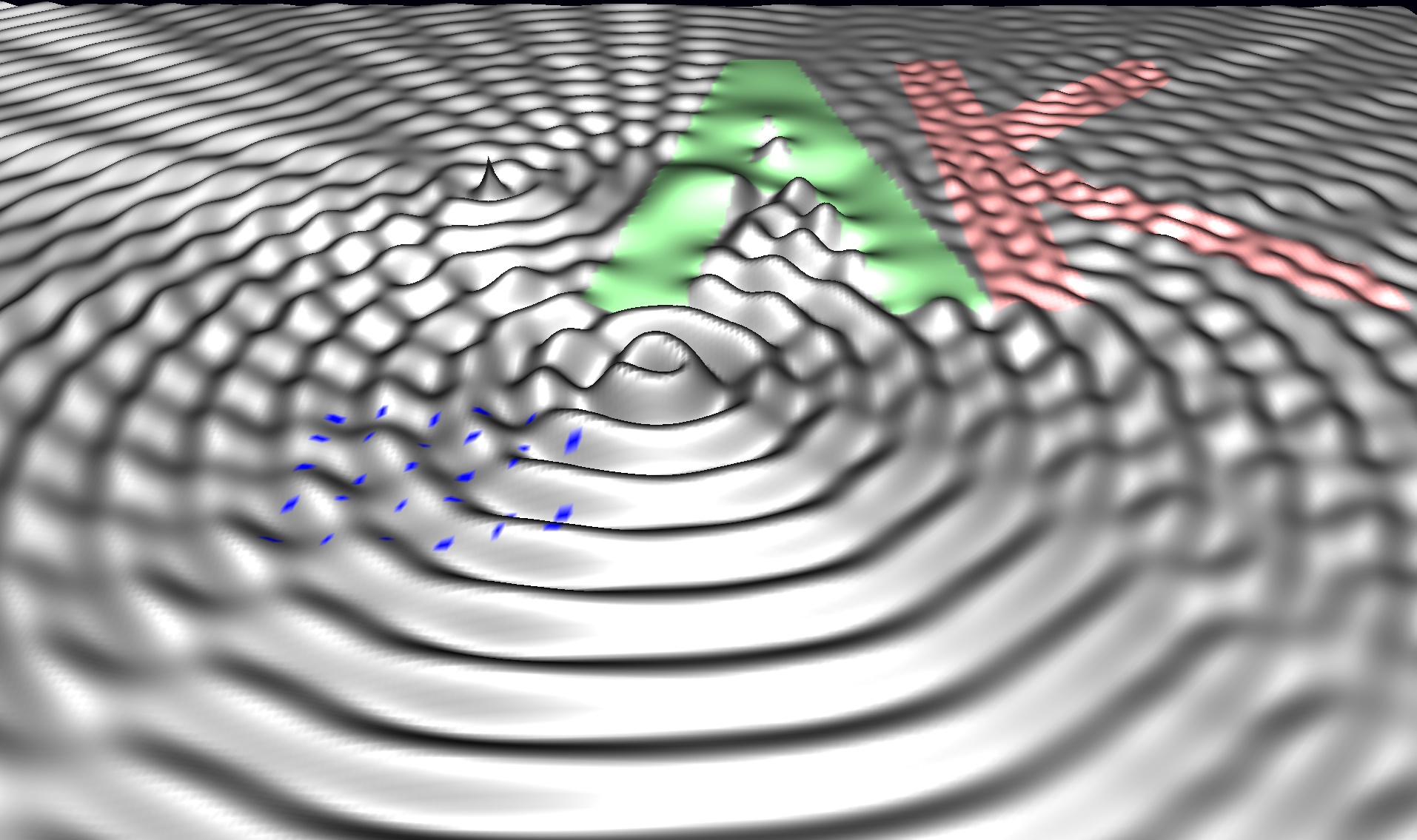}} \\ (a)
\end{minipage}
\hfill
\begin{minipage}[h]{0.49\linewidth}
\center{\includegraphics[width=1\linewidth]{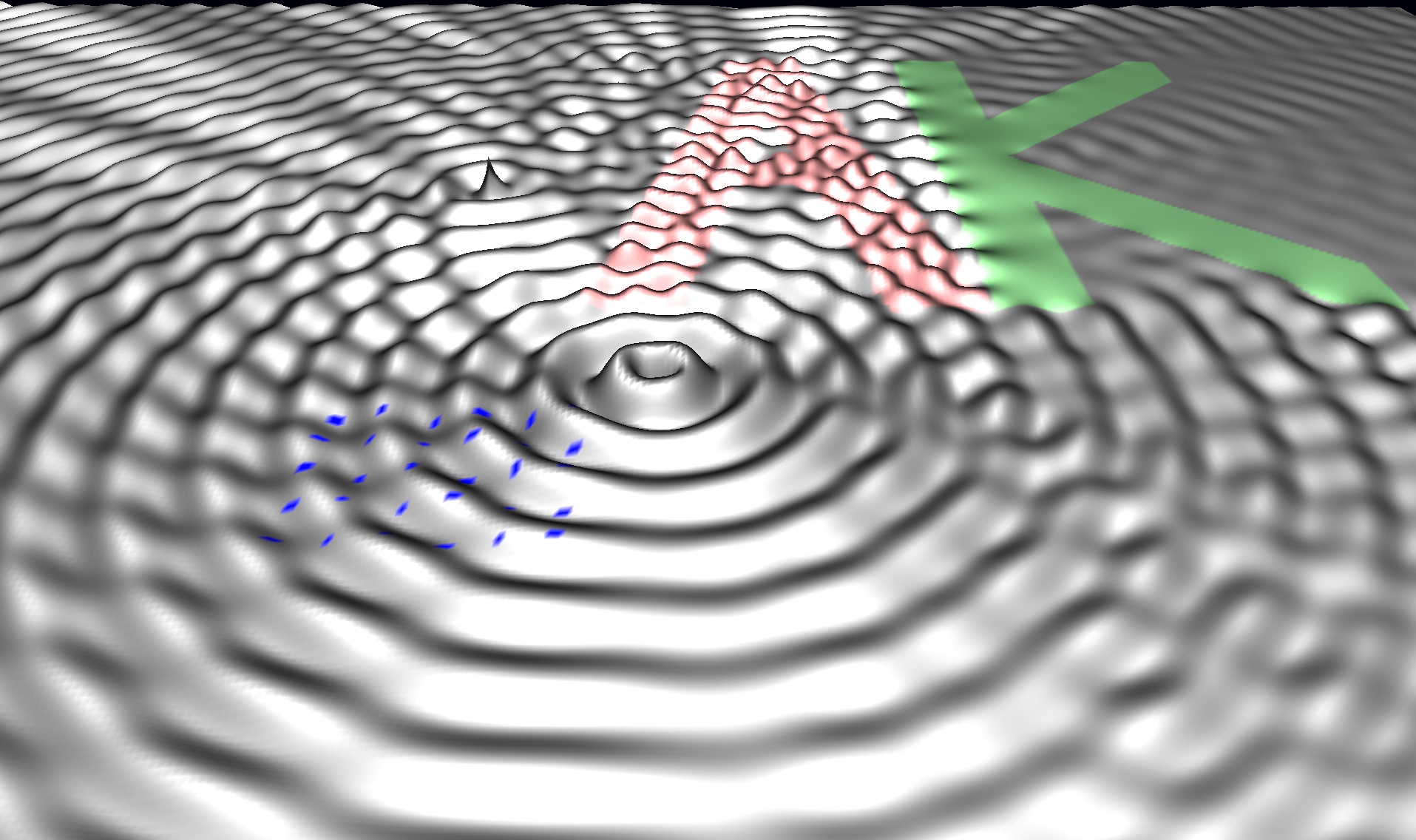}} \\ (b)
\end{minipage}}
\caption{(color online) Wave fields in the uniform medium with
defects and sources (computed by \er{102}), where the red defect
corresponds to a slow material and the green defect is fast. Inverse
problem consists of the recovery of defect properties from the
information about amplitudes of waves observed at the detectors
(blue points).} \label{fig1}
\end{figure}

The current research is inspired by \cite{RU,RU1}, where the authors
considered the problem of recovering smooth compactly supported
potential $q$ in the continuous equation of Schrodinger type
$U_{tt}-\D U+qU=F$ from its far backscattering data. Some
interesting observations devoted to inverse scattering problems on
discrete periodic graphs are given in \cite{IK}, \cite{A}. Note also
that for continuous media with sparse distributed point scatters
there are stable and efficient methods of recovery of scattering
properties (see \cite{CMP}, \cite{CSV}), especially if the multiple
scattering is negligible. In the present paper, we consider the
discrete wave equation
\[\lb{000a}
 S^2_{\bf n}\frac{\pa^2U_{\bf n}}{\pa t^2}-\D_{\rm
discr} U_{\bf n}=F_{\bf n},\ \ {\bf n}\in\Z^d
\]
($d$ is the dimension) and try to recover the slownesses $S_{\bf n}$
from the information about the amplitudes of waves observed at some
nearby points $\cR\ss\Z^d$, see examples in Fig. \ref{fig1}. The
discrete Laplacian in \er{000a} is
\[\lb{000b}
 \D_{\rm discr} U_{\bf n}=\sum_{{\bf n}'\sim{\bf n}}(U_{{\bf n}'}-U_{\bf
 n}),
\]
where $\sim$ means neighboring points (we have $2d$ neighboring
points). We assume that the slowness $S_{\bf n}^2=s^2$ is uniform at
each point of the lattice $\Z^d$ except some defect points $\cN$,
where
\[\lb{deffslow}
 S_{{\bf n}}^2=s^2+s_{\bf n}^2,\ \ {\bf n}\in\cN.
\]
The term $s_{\bf n}$ is a so-called defect perturbation of the
constant slowness $s$. We assume also that the sources have the form
\[\lb{sources}
 F_{\bf n}=\sum_{j=1}^{M}e^{-i\o_{j} t}\sum_{{\bf m}\in\cF_j}F^{j}_{{\bf m}}\d_{{\bf m}{\bf
 n}},
\]
where $\d$ is the Kronecker delta. The set $\cF_j\ss\Z^d$ consists
of locations of sources of the same frequency $\o_j$, the constant
amplitudes $F^{j}_{{\bf m}}$ are all non-zero. It is natural to
assume that all frequencies are different $\o_i\ne\o_j$, $i\ne j$.
Note that it is possible to have many frequencies at one point since
$\cF_i\cap\cF_j$ can be non-empty for $i\ne j$. The number of
defects $N=\#\cN$, receivers $R=\#\cR$, and different frequencies
$M$ are finite numbers. For simplicity, we assume that all $\o_j^2$
do not belong to the spectrum $[0,4d]$. As it is shown in \cite{K}
the solution of the equation \er{000a} has the form
\[\lb{001a}
 U_{\bf n}=\sum_{j=1}^MU_{{\bf n}}^je^{-i\o_{j}t},\ \ {\bf
 n}\in\Z^d,
\]
where the constant amplitudes $U_{\bf n}^j$ can be explicitly
expressed as rational functions of defect perturbations (see
\er{100}, \er{102}, and details in \cite{K}). We will focus on the
inverse problem. Suppose that we record amplitudes at the receivers
$\cR$. Thus we know the vectors
\[\lb{002a}
 {\bf u}_j=(U_{{\bf r}_i}^j)_{i=1}^R\ \ for\ \
 j=1,...,M,
\]
where $\cR=\{{\bf r}_i\}_{i=1}^R$. Suppose that we know the
approximation location of the defects. For convenience we can assume
that the set $\cN$ is known, and it is a sufficiently large set
which cover all defect points. If some point ${\bf n}$ of this large
set $\cN$ is non-defect point then $s_{\bf n}=0$. We assume also
that the information about the sources is available. All these data
will be used to determine the unknown defect perturbations $s_{\bf
n}$, ${\bf n}\in\cN$. Let ${\bf k}=(k_i)\in[-\pi,\pi]^d$ and
introduce
\[\lb{006a}
 A_j=2d-(\o_j s)^2-2\sum_{i=1}^d\cos k_i,\ \ \
 \langle...\rangle=\frac1{(2\pi)^d}\int\limits_{[-\pi,\pi]^d}... d{\bf
 k},
\]
\[\lb{006}
{\bf a}=\ma e^{-i{\bf n}_1\cdot{\bf k}} \\
... \\ e^{-i{\bf n}_N\cdot{\bf k}}\am,\ \ \ {\bf a}_{\bf
m}^j=\lt\langle\frac{{\bf a}e^{i{\bf m}\cdot{\bf
k}}}{A_j}\rt\rangle,\ \ \ {\bf s}=(s^2_{{\bf n}_i})_{i=1}^N,\ \ \
 {\bf S}=\diag({\bf s}),
\]
where $\cN=\{{\bf n}_i\}_{i=1}^N$ and $\diag({\bf vector})$ denotes
a diagonal matrix with components of the {\bf vector} on the main
diagonal. Introduce also
\[\lb{007}
{\bf c}=\ma e^{-i{\bf r}_1\cdot{\bf k}} \\
... \\ e^{-i{\bf r}_R\cdot{\bf k}}\am,\ \ \ {\bf c}_{\bf
m}^j=\lt\langle\frac{{\bf c}e^{i{\bf m}\cdot{\bf
k}}}{A_j}\rt\rangle,\ \ \ {\bf C}_j=\lt\langle\frac{{\bf c}{\bf
a}^*}{A_j}\rt\rangle,\ \ \ {\bf A}_j=\lt\langle\frac{{\bf a}{\bf
a}^*}{A_j}\rt\rangle,
\]
where $ ^*$ denotes Hermitian conjugation. Let us introduce the sets
of so-called {\it admissible} slownesses ${\bf s}$:
\[\lb{007a}
 \cG_{\rm adm}=\bigcap_{j=1}^M\cG_{\rm adm}^j,\ \ where\ \ \cG_{\rm adm}^j=\{{\bf s}:\ \det{\bf
 G}_j\ne0\},\ \ {\bf G}_j={\bf I}-\o^2_j{\bf A}_j{\bf S}.
\]
For {\it admissible} defects the direct problem of determining the
amplitudes $U_{\bf n}$ which satisfy \er{000a} has a unique
solution. Almost all defects encountered in the nature are {\it
admissible}. Only the structures with very unusual properties can
contain {\it non-admissible} defects. So the considering of {\it
admissible} defects is probably not a significant limitation. The
next theorem describes analytically the set of all possible defects
${\bf s}$ which correspond to given amplitudes ${\bf u}_j$ of waves
measured at the receivers.

\begin{theorem}\lb{T1} If the amplitudes ${\bf u}_j$, $j=1,...,M$ at the receivers
correspond to some defect ${\bf s}$ then there are ${\bf
x}_j\in\C^N$ such that
\[\lb{008}
 {\bf C}_j{\bf x}_j={\bf u}_j-\sum_{{\bf m}\in\cF_j}{\bf c}_{\bf m}^jF^j_{\bf m},\ \ j=1,...,M
\]
and ${\bf s}\in\cS=\cap_{j=1}^M\cS_j$, where
\[\lb{009}
 \cS_j=\{\o_j^{-2}({\bf x}+{\bf x}_j)/({\bf A}_j{\bf x}+{\bf A}_j{\bf x}_j+\sum_{{\bf m}\in\cF_j}{\bf a}_{\bf m}^jF_{\bf m}^j):\ {\bf x}\in{\rm ker}{\bf
 C}_j\}
\]
(ratio $/$ means the component-wise ratio of two vectors). Moreover,
for any defect ${\bf s}\in\cS\cap\cG_{\rm adm}$ the amplitudes at
the receivers are ${\bf u}_j$, $j=1,...,M$.
\end{theorem}

{\bf Remark.} The case of zero denominator in \er{009} needs an
additional analysis. Roughly speaking, for the component with zero
both numerator and denominator we can take any value, and the
manifold $\cS_j$ consists of the main (regular) part of the
dimension as ${\rm ker}{\bf C}_j$ (see also 5) below) and linear
subspaces corresponding to $0/0$.

{\bf Comments on applications.} 1) If ${\rm ker}{\bf C}_j=\{{\bf
0}\}$ for some $j$ then we can uniquely recover the defect ${\bf s}$
from the information about amplitudes of waves. This can happen when
the location of defects $\cN$ is a moderately sparse set. But for
most applications the defect area is large and dense, e.g. it is a
square $\cN=[a_i,b_i]^d$ with sufficiently large $b_i-a_i$, because
we want to cover all possible defects. In this case, the matrices
${\bf C}_j$ have non-trivial kernels (see \cite{K}). Also each
$\cS_j\ss\C^N$ is a manifold with dimension $N-\rank{\bf C}_j$. Thus
if the number of different frequencies is greater than $N$ then we
can expect that $\cS$ consists of a single point, since the
intersection of manifolds is usually a manifold of a smaller
dimension. If $\cS$ consists of a unique point then the inverse
problem is solved uniquely. This is important for practical
applications where the goal is a unique solution of inverse
problems. In the case where $\cS$ consists of more than one point,
we can increase the chance of recovering the defect by using some
available additional information; for example, if we know that all
slownesses of defective points are positive and bounded $(\le B)$.
Then the set $[0,B]^N\cap\cS$ is not large and it can provide useful
characteristics of the defect
because usually ${\bf x}_j\not\in{\rm ker}{\bf C}_j$ and $\cS$
consists of complex vectors that are located far from the origin
${\bf 0}$.

\begin{figure}[h]
{\begin{minipage}[h]{0.49\linewidth}
\center{\includegraphics[width=1\linewidth]{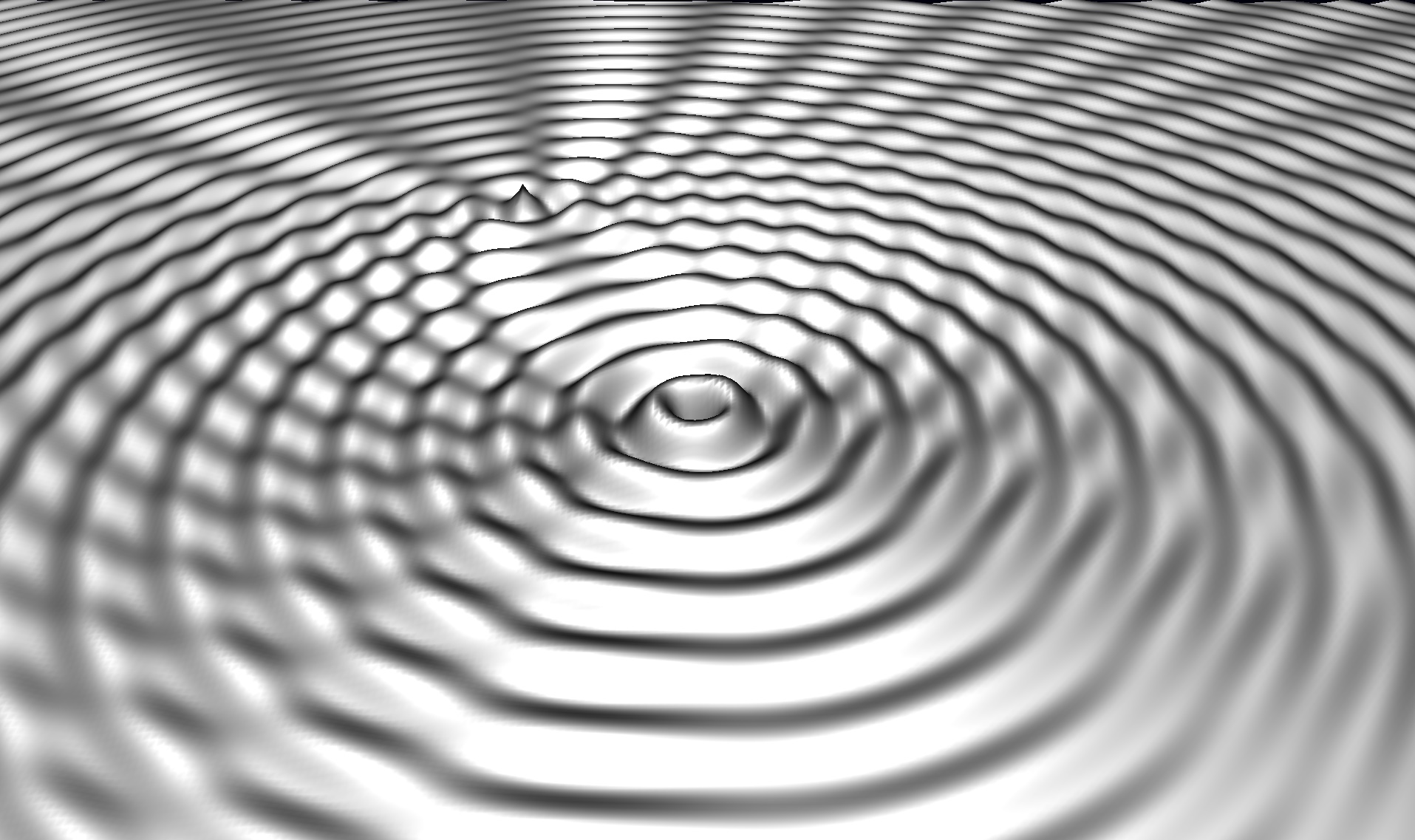}} \\ (a)
\end{minipage}
\hfill
\begin{minipage}[h]{0.49\linewidth}
\center{\includegraphics[width=1\linewidth]{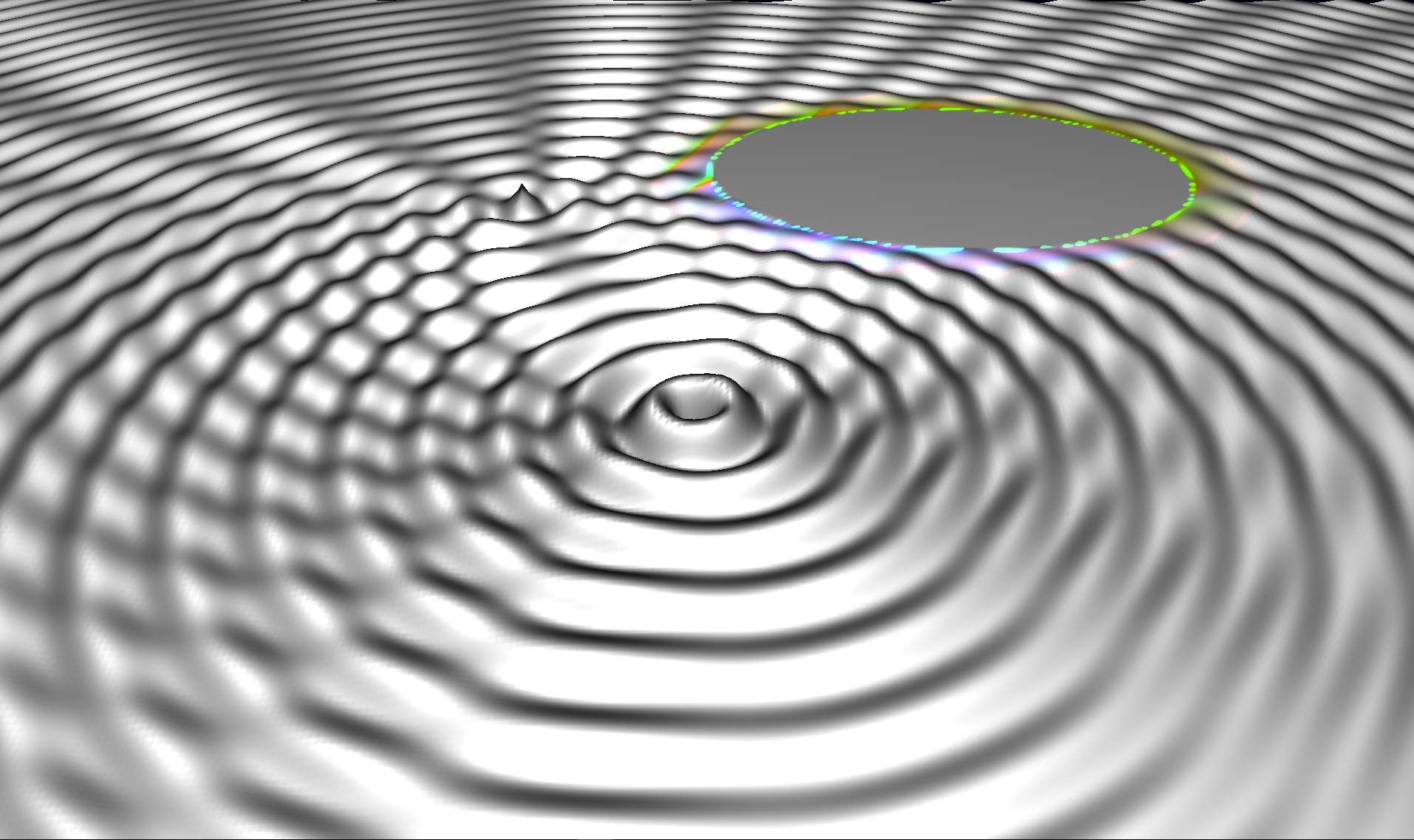}} \\ (b)
\end{minipage}}
\caption{(color online) Wave fields in the uniform medium without
defects (a) and with cloaking defect (b).} \label{fig2}
\end{figure}

2) In this paragraph we assume that the configuration of sources is
fixed. We assume also that the location $\cN$ of defect is known.
Suppose we want to construct the defect that cannot be detected at
the receivers. For this we only need to take the unperturbed field
${\bf u}_j=\sum_{{\bf m}\in\cF_j}{\bf c}_{\bf m}^jF^j_{\bf m}$ in
Theorem \ref{T1} which leads to the sets $\cS_j^0$ \er{009} with
${\bf x}_j=0$. The intersection $\cS^0=\cap_{j=1}^M\cS^0_j$ is
precisely the set of all possible invisible defects corresponding to
given receivers. A defect which is invisible to any locations of
external receivers is called a cloaking device. Roughly speaking, we
cannot detect this "global" defect from the outside, see Fig.
\ref{fig2}. The set of all possible cloaking devices is now the
intersection of all of $\cS^0$ corresponding to all configurations
of receivers (enough to take a finite ring of receivers around the
defect). By the same arguments as in the first paragraph, it may
happen that for a large number $M$ of frequencies this set is empty.
Nevertheless for monochromatic sources ($M=1$), the cloaking device
can be constructed explicitly together with an arbitrary wave field
near the device, see also \cite{K}. In particular, we can assume
that the amplitudes of wave field is $0$ inside the cloaking device
(cloaking insulator). In this case, any object can be hidden inside
the region of zero wave amplitudes and is not detectable from the
outside. To sum up, taking ${\bf u}_j={\bf 0}$ and/or ${\bf
u}_j=\sum_{{\bf m}\in\cF_j}{\bf c}_{\bf m}^jF^j_{\bf m}$, and/or any
other amplitudes we can construct cloaking devices with various
properties. Some additional information about the applications of
cloaking devices is provided in \cite{Maldovan,NS,TD}.

3) Almost all the matrices introduced above consist of the elements
of the form
\[\lb{010}
 a_{{\bf n}}^j=\langle A_j^{-1}e^{i{\bf n}\cdot{\bf k}}\rangle.
\]
These components are symmetric in ${\bf n}$, i.e. if we change the
sign of any entry $n_i$ of ${\bf n}$ then $a_{{\bf n}}^j$ does not
change. They also satisfy the following identity
\[\lb{011}
 \sum_{{\bf n}'\sim{\bf n}} a^j_{{\bf n}'}=(2d-\o_j^2s^2)a^j_{{\bf
 n}}+\d_{{\bf n}{\bf 0}}.
\]
This equation explains why ${\rm ker}{\bf C}_j$ (and hence $\cS_j$)
are non-trivial for dense sets of defects, since $a^j_{{\bf n}}$ is
completely determined by $a^j_{{\bf n}'}$ where ${\bf n}'$ are
neighboring points. By the same reason it will be more effective to
use sparse sets of receivers.

4) We have considered the frequencies $\o_j$ that do not belong to
the spectrum $[0,4d]$ (passband). They can be complex frequencies
$\o_j=\a_j-i\b_j$ (which means that we have a source attenuation
factor $e^{-\b_jt}$), or they can be high real frequencies. At the
same time there are many problems where $\o_j\in[0,4d]$; for
example, the small real frequencies appear in the problems of long
wave propagation. For such frequencies $A_j^{-1}$ are not
well-definite but we can consider the limit $\o_j-i\ve$ with
$\o_j\in[0,4d]$ and $\ve\to+0$. Except for some specific cases of
frequencies ($\o=0$, $\o=4d$) it is not difficult to show that the
limit of $a_{{\bf n}}^j$ exists and hence we can extend our results
to the frequencies from the passband. Such examples are considered
in \cite{K}, where some computational aspects for $a^j_{\bf n}$ with
$\ve\to+0$ are also discussed.

5) Denote the ratio of two vectors in \er{009} as ${\bf s}({\bf
x})$, where ${\bf x}\in{\rm ker}({\bf C}_j)$. By \er{106} we have
that the element ${\bf y}_j$ from ${\rm ker}({\bf C}_j)$ satisfying
\er{104} is uniquely defined. In other words if ${\bf s}({\bf
x}_1)={\bf s}({\bf x}_2)\in\cS_j\cap\cG_{\rm adm}^j$ then ${\bf
x}_1={\bf x}_2$, and hence the mapping ${\bf s}$ is a
parametrization of the manifold.

\section{Proof of Theorem \ref{T1}} \lb{sect1}

Taking Fourier series (see details in \cite{K})
\[\lb{100}
 u_j=\sum_{{\bf n}\in\Z^d}U_{\bf n}^je^{i{\bf n}\cdot{\bf k}}
\]
we can equivalently rewrite the problem \er{000a} as a set of
integral equations
\[\lb{101}
 A_ju_j-\o_j^2{\bf a}^*{\bf S}\langle{\bf a}u_j\rangle=\sum_{{\bf m}\in\cF_j}F^j_{\bf m}e^{i{\bf m}\cdot{\bf
 k}},\ \ j=1,...,M.
\]
If ${\bf s}$ is {\it admissible} then there is a unique solution of
\er{101}:
\[\lb{102}
 u_j=A_j^{-1}\sum_{{\bf m}\in\mF_j}F_{\bf m}^j(\o_j^2{\bf a}^*{\bf S}{\bf G}_j^{-1}{\bf a}_{\bf m}^j+e^{i{\bf m}\cdot{\bf
 k}}).
\]
Multiplying \er{101} by $A_j^{-1}{\bf c}$ and taking the integral
$\langle...\rangle$ we obtain
\[\lb{103}
 {\bf u}_j-\o_j^2{\bf C}_j{\bf S}\langle{\bf a}u_j\rangle=\sum_{{\bf m}\in\cF_j}F_{\bf m}^j{\bf
 c}_{\bf m}^j
\]
which means that there exist ${\bf x}_j=\o_j^2{\bf S}\langle{\bf
a}u_j\rangle$ satisfying \er{008}. This means also that
\[\lb{107}
 {\bf S}\langle{\bf a}u_j\rangle=\o_j^{-2}({\bf y}_j+{\bf x}_j),
\]
where ${\bf y}_j\in{\rm ker}{\bf C}_j$. Multiplying \er{101} by
$A_j^{-1}{\bf a}$, taking the integral $\langle...\rangle$, and
substituting \er{107} into \er{101} we obtain that
\[\lb{108}
 \langle{\bf a}u_j\rangle={\bf A}_j{\bf y}_j+{\bf A}_j{\bf
 x}_j+\sum_{{\bf m}\in\cF_j}{\bf a}_{\bf m}^jF_{\bf m}^j.
\]
Equations \er{107} and \er{108} mean that ${\bf s}$ belongs to
$\cS$.

Now, suppose that we have some ${\bf x}_j$ satisfying \er{008}. We
take some ${\bf s}\in\cS\cap\cG_{\rm adm}$. Then there are ${\bf
y}_j\in{\rm ker}{\bf C}_j$ such that
\[\lb{104}
 {\bf s}=\o_j^{-2}({\bf y}_j+{\bf x}_j)/({\bf A}_j{\bf y}_j+{\bf A}_j{\bf
 x}_j+\sum_{{\bf m}\in\cF_j}{\bf a}_{\bf m}^jF_{\bf m}^j)
\]
or (because ${\bf S}=\diag({\bf s})$)
\[\lb{105}
 {\bf y}_j+{\bf x}_j=\o_j^{2}{\bf S}({\bf A}_j{\bf y}_j+{\bf A}_j{\bf
 x}_j+\sum_{{\bf m}\in\cF_j}{\bf a}_{\bf m}^jF_{\bf m}^j)
\]
which leads to
\[\lb{106}
 {\bf y}_j+{\bf x}_j=\o_j^{2}{\bf S}{\bf G}_j^{-1}\sum_{{\bf m}\in\cF_j}{\bf a}_{\bf m}^jF_{\bf m}^j.
\]
Note that in the implication $\er{105}\Rightarrow\er{106}$ we use
the following well known fact: {\it Let ${\bf U},{\bf V}$ be two
arbitrary square matrices of the same size. If ${\bf I}-{\bf U}{\bf
V}$ is invertible then ${\bf I}-{\bf V}{\bf U}$ is also invertible
and
$$
 ({\bf I}-{\bf U}{\bf V})^{-1}{\bf U}={\bf U}({\bf I}-{\bf V}{\bf
 U})^{-1},
$$
where ${\bf I}$ denotes the identity matrix.} Consider \er{102}
(which is the unique solution of \er{101}) with ${\bf S}=\diag({\bf
s})$. Multiplying \er{102} by ${\bf c}$, taking the integral
$\langle...\rangle$, and using \er{106} we obtain that
\[\lb{109}
 \langle{\bf c}u_j\rangle={\bf C}_j({\bf y}_j+{\bf x}_j)+\sum_{{\bf m}\in\cF_j}{\bf
 c}_{\bf m}^jF_{\bf m}^j={\bf u}_j,
\]
where we also use ${\bf y}_j\in{\rm ker}{\bf C}_j$ and \er{008}.
Equation \er{109} means exactly that given ${\bf s}$ corresponds to
the amplitudes ${\bf u}_j$ at the receivers, see \er{100} and
\er{002a}.

\section*{Acknowledgements}
This work was partially supported by the RSF project
N\textsuperscript{\underline{o}}15-11-30007. I would also like to
thank Prof. Daphne J. Gilbert for useful discussions.

\bibliography{invfreq}

\end{document}